# Do Social Bots Dream of Electric Sheep? A Categorisation of Social Media Bot Accounts

**Stefan Stieglitz**
Department of Computer Science and Applied Cognitive Science
University of Duisburg-Essen
Duisburg, Germany
Email: stefan.stieglitz@uni-due.de

**Florian Brachten**
Department of Computer Science and Applied Cognitive Science
University of Duisburg-Essen
Duisburg, Germany
Email: florian.brachten@uni-due.de

**Björn Ross**
Department of Computer Science and Applied Cognitive Science
University of Duisburg-Essen
Duisburg, Germany
Email: bjoern.ross@uni-due.de

**Anna-Katharina Jung**
Department of Computer Science and Applied Cognitive Science
University of Duisburg-Essen
Duisburg, Germany
Email: anna-katharina.jung@uni-due.de

## Abstract

So-called 'social bots' have garnered a lot of attention lately. Previous research showed that they attempted to influence political events such as the Brexit referendum and the US presidential elections. It remains, however, somewhat unclear what exactly can be understood by the term 'social bot'. This paper addresses the need to better understand the intentions of bots on social media and to develop a shared understanding of how 'social' bots differ from other types of bots. We thus describe a systematic review of publications that researched bot accounts on social media. Based on the results of this literature review, we propose a scheme for categorising bot accounts on social media sites. Our scheme groups bot accounts by two dimensions – Imitation of human behaviour and Intent.

## 1　Introduction

Social media permeate society. Brands use them to influence customers' purchase intentions (Xie and Lee 2015) and political candidates use them to disseminate information to their supporters, but their unregulated nature has given rise to a flood of information of questionable credibility (Wattal et al. 2010). It has been shown that businesses such as hotels are posting manipulated content on social media to gain an unfair advantage over their competitors (Mayzlin et al. 2014). Against this backdrop, it becomes clear that social media have increasingly become interesting for people and organisations looking to influence the discussion on a certain topic. The automated dissemination of messages promises to be an efficient way to reach many people with little effort. The reasons for spreading automated content range from the dissemination of information (e.g. stock prices, weather data), spam, malware or advertisement to political intentions (Alarifi et al. 2016). Recently social bots, algorithms programmed to mimic human behaviour on social media platforms, have become increasingly attractive for people and organisations aiming to automatically distribute their messages to many recipients at very low costs. Current studies reveal that social bots are involved in online discussions about current political events, such as the armed conflict between Ukraine and Russia and the war in Syria by spamming the discussion with one-sided arguments or unrelated content to distract participants (Abokhodair et al. 2015; Hegelich and Janetzko 2016). The mere presence of automated actors in vital opinion-shaping discussions provokes the fear of manipulation and thus ethical concerns. This has led to increasing press coverage on the expected influence of social bots (e.g. Dewey 2016; Fuchs 2016; Guilbeault and Woolley 2016; Lobe 2016). The great public interest in social bots underlines the importance of a profound scientific analysis of the topic.

As the topic of social bots is still young, and as it is approached from multiple angles, the terms and definitions used to describe related phenomena are diverse. Sometimes several different terms are used to label the same concept, and sometimes a single term such as social bots is used to describe different things. This leads to a diffuse use of terms and subsequently to imprecise theoretical foundations in this area. For example, some researchers use the term "social bot" for any account on social media run by an algorithm (e.g. Forelle et al. 2015), while others use a much more restrictive definition, for example as "computer programs designed to use social networks by simulating how humans communicate and interact with each other" (Abokhodair et al. 2015, p.25). Some researchers consider a social bot a potentially harmful adversary by definition (Boshmaf et al. 2013). This confusion around terms and definitions means that there is a clear need for a structured approach to the topic of social bots.

At the same time, the prevalence of bots on social media raises interesting research questions and challenges for the IS research community. Identification techniques, communication patterns and the impact of social bots on individuals and companies are only three examples of possible research topics in IS. Our literature review shows that there has been very little IS research into these topics so far. Given that this topic is clearly highly relevant for the IS community, there is a noticeable research gap. To begin addressing this gap and provide guidance for future research, it is first necessary to clearly delineate the types and activities of bots on social media.

This paper summarises the types of bot accounts active in social media. Moreover, we discuss definitions and terms that are used in academic articles. In order to structure different types of social bots, we develop a categorisation scheme that builds upon our findings from literature. The literature review will (a) contribute to the specification of the field and will offer an excellent starting point for researchers who plan to start investigating bots on social media, (b) allow new forms of bots to be assessed faster through a categorisation system, as it offers a scheme to group and classify them and (c) provide a clear definition of social bots which demarcates them from other forms of automated actors in social media.

## 2　Background

This article concerns bots that run or control social media accounts. We do not consider bots that make use of social media features but do not control their own accounts, e.g. botnets that communicate by surreptitiously injecting messages into photos uploaded by the user (Nagaraja et al. 2011).

Even within this relatively narrow field of research that concentrates on bots in the context of social media, there is an enormous diversity of bots. Moreover, researchers from different backgrounds tend to approach bots from different angles using various theories and concepts. Information security researchers view bots in an adversarial role, and demonstrate the feasibility of hypothetical attacks or devise potential defence mechanisms. For example, Pantic and Husain who research botnets that communicate by actively controlling their own social media accounts state that "Botnet software is a type of malicious software (malware) that is most often placed on a victim's computer silently" (2015, p. 172). Researchers in journalism explore how useful news reporting bots are transforming their field (Lokot and Diakopoulos 2016). Social scientists may place their own bots to explore how humans react to them (Wilkie et al. 2015). It becomes obvious that researchers in different disciplines develop their own unique perspectives and theoretical foundations.

The resulting confusion extends to the terminology, which is equally diverse. For example, a large number of the papers concerning bots on Twitter (or Twitter bots) address political goals and consider social bots in this context, e.g. during the Syrian war (Abokhodair et al. 2015), the crisis in Ukraine (Hegelich and Janetzko 2016), Venezuelan politics (Forelle et al. 2015) and regional elections in Germany (Brachten et al. 2017). While these authors are all interested in examining how bots can be identified and the extent to which they are used in practice, there seems to be disagreement in naming the researched aspects. Some authors use similar definitions of social bots: Abokhodair et al. define them as "software designed to act in ways that are similar to how a person would act in the social space" (2015, p. 840), and Hegelich and Janetzko call them "automatic programs [that] are mimicking humans" (2016, p. 579). Forelle et al. describe social bots as "computer-generated programs that post, tweet, or message of their own accord" (2015, p.1). While the first two sources both point out that those bots imitate humans, this aspect is missing in the third citation. Boshmaf et al. (2013) mention a further component: According to them, a social bot "is an automation software that controls an adversary-owned or hijacked account on a particular OSN, and has the ability to perform basic activities such as posting a message and sending a connection request" (p. 556). Here again the aspect of imitating a human user is missing, but the authors mention hijacked accounts, which were not prevalent in the other definitions. Another important difference is, again, the reference to adversaries implying that social bots are, by definition, opponents. This example demonstrates how terminological ambiguity and the lack of a shared conceptual understanding go hand in hand.

Aside from the lacking consensus on what exactly is to be understood by the term "social bots", several papers use other labels to describe phenomena which could be understood as social bots according to some of the above definitions. Igawa et al. (2016) write, that "[o]n Twitter social robots, called 'bots', pretend to be human beings in order to gain followers and replies from target users and promotes a product or agenda" (p. 73). The definition describes features of social bots (respectively goals of the bot developer) simply calling the relevant accounts "bots". The same goes for Larsson and Moe (2015) who define bots as "a piece of more or less automated computer software, programmed to mimic the behaviour of human Internet users" (p. 362). While their definition is more general as it does not limit the accounts to Twitter, it also features the aspect of human imitation.

These examples show that a broad consensus is missing but needed to precisely describe the researched aspects. In order to reach this goal, a comprehensive overview over the relevant literature is needed. The exact procedure is described in the next chapter.

## 3  Method

The following literature review was conducted based on the systematic process proposed by vom Brocke et al. (2009), which includes the taxonomy for literature reviews by Cooper (1988). Primarily the scope of the research was limited to bots that run or control social media accounts regardless of their intentions and methods. Information Systems (IS) researchers and general scholars have been defined as the main audience of the literature review.

To gather relevant literature, we first conducted a search in three databases: Scopus, ScienceDirect and the AIS Electronic Library (AISeL). While Scopus and ScienceDirect allow us to identify research articles on a broader level, AISeL is a source explicitly used by academics from IS. First, we searched for literature that either included the term bot or socialbot and one of the terms Twitter, Facebook, "social networks" or "social media" in the title, abstract or keywords. To get a systematic overview of recent high-quality academic publications, we limited the search to peer-reviewed articles that were released in or after 2007. Due to our own language skills and comparability we only consider articles that were written in English. To ensure the scientific quality of the publications, we only considered peer-reviewed papers from conference proceedings and scientific journals.

The search was carried out on 31 July, 2017. We found 187 entries in the Scopus database and 19 in ScienceDirect. The search in AISeL led to no results. Within the 187 Scopus entries we found three duplicates (due to different titles but identical contents) which were excluded. We then matched the entries from Scopus and ScienceDirect to filter out duplicates found in both databases, which led to the exclusion of twelve entries.

As a next step the titles and abstracts of the remaining 191 entries were examined in more detail to assess if they were relevant for our main goal of observation. Of the 191 entries, 68 were excluded unanimously from further investigation due to irrelevance (e.g. one paper that dealt with search engine optimisation used the term bot in reference to the Google crawler and the term social media in its keywords). Of the remaining entries, 88 were included unanimously while for 35 entries the authors' evaluation differed. Those entries were re-evaluated in a group discussion. This discussion led to 16 out of the 35 entries being included, forming a sample of 104 papers. Since one paper could not be obtained, the final sample consisted of 103 papers.

Figure 1 shows the number of papers published by year on the topic of bot accounts in social media. As the figure shows, there is a fairly consistent increase in the number of papers published. The year 2014 can be described as an outlier. However, closer inspection did not reveal why the number of publications dropped in this year. For 2017, only the first seven months (January to July) are present, and the number is therefore lower than the peak in 2016. Also, the final sample consisted only of papers released in or after 2010 as no papers were identified between 2007–2009.

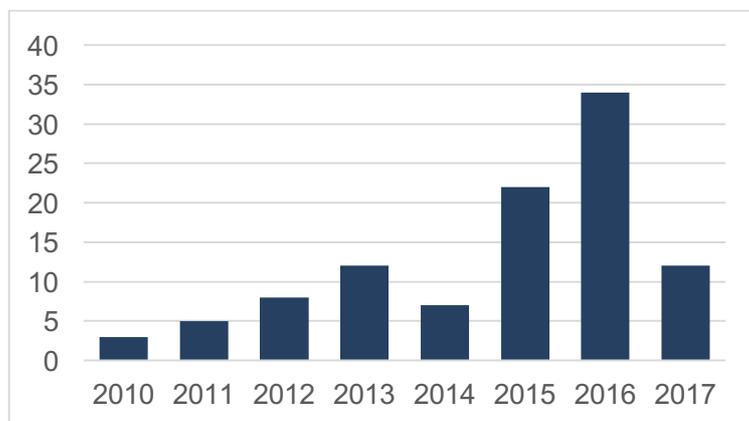

*Figure 1: Number of papers published in each year in the examined sample*

## 4    A categorisation of social media bot accounts

How can the bots active on social media be distinguished from one another? Two distinctions are commonly made in the literature.
1. First, bots are distinguished into benign and malicious bots (Ferrara et al. 2016). Benign bots aggregate content, respond automatically, and perform other useful services. Malicious bots, in contrast, are designed with a purpose to harm. Our analysis of the literature shows that the categorisation of social bots into malicious and benign ones is widely accepted.
2. The second distinction can be traced to Boshmaf et al. (2013), who wrote that the crucial difference between social bots and other types of bots is that the former "is designed to pass itself off as a human being. This is achieved by either simply mimicking the actions of a real OSN user or by simulating such a user using artificial intelligence" (p. 556).

In the following, we describe these categories in more detail, show how they were applied in related papers, and give specific examples. We also show that these distinctions are orthogonal to each other, thus giving rise to a two-dimensional categorisation of bots.

### 4.1    Intentions – malicious or benign?

Due to the multiple research perspectives to study (social) bots, the literature offers a wide range of definitions, terms and classifications. The most common distinction is that between benign and malicious bots (Ferrara et al. 2016). 52 out of the 103 papers used the term malicious to classify certain bot accounts or describe their behaviour (e.g. Boshmaf et al. 2016; Chu et al. 2012; Edwards et al. 2014; He et al. 2017). For example, Freitas et al. point out that "Socialbots can have many applications, with good or malicious objectives" (2015, p. 25). Bots which spread spam or falsehoods are generally defined as malicious (Bhat and Abulaish 2013; Lokot and Diakopoulos 2016; Main and Shekokhar 2015). Bots delivering useful automated information such as news or weather reports, are usually labelled *good*, *helpful* or *benign* (Alarifi et al. 2016; Chu et al. 2012). Here, 20 papers use the latter term to label such bots.

The most frequently mentioned benign bots active in social media are those used by mass media, grassroot journalists and bloggers to automatically post recently published articles or breaking news. Lokot and Diakopoulos (2016) classify these news bots based on their input and sources, outputs, algorithms and intent or function. Weather bots, sport bots, traffic bots, niche news bots and geo-specific bots are only some examples of the news bots they identified through their research. Bots can contribute positively to the recruitment of volunteers (Flores-Saviaga et al. 2016; Savage et al. 2016). Flores-Saviaga et al. (2016) deployed two bots which contacted and motivated experts to mention women who were still missing an entry on Wikipedia. Furthermore Tsvetkova et al. (2017) report that recruitment bots, as well editing bots and anti-vandalism bots are frequently used on Wikipedia. Another type of benign bots is chat bots, which can be used by enterprises to limit the need for human involvement in business-to-customer communication. In addition to that, chat bots can be used to respond to customer questions during events (Salto Martínez and Jacques García 2012).

However, the ill intents of bots are much more diverse. They include: spam, the theft of personal data and identities, the spreading of misinformation and noise during debates, the infiltration of companies and the diffusion of malware (Abokhodair et al. 2015; Bhat and Abulaish 2013; Bokobza et al. 2015; Elyashar et al. 2015; Goga et al. 2015; Zhang et al. 2013; Zhu et al. 2013). Previous literature has introduced specific terms to describe bots involved in certain malicious acts. *Commercial bots*, according to Subrahmanian et al. (2016), include e.g. spam bots, and pay bots "which copy content from respected sources and paste it into micro URLs that pay the bot creator for direct traffic to that site" (p. 38). Bots which operate on social media with a fake identity or have the aim of impersonation are often described as *sybils* – 18% of the examined papers used that term (e.g. Goga et al. 2015; Paradise et al. 2014). According to Goga et al. (2015), the three main types of identity attacks are celebrity impersonation attacks (duplication of a celebrity account by a sybil account), social engineering attacks (which aim at motivating friends/followers to disclose private data) or doppelgänger bot attacks (which are copies of user accounts, in order to use their identity to slip through the networks detection systems). However, the term *sybil* is not clearly distinguishable from the term *bot*, as bots operate with fake identities, too. Therefore, the terms *sybil* and *bot* are often used interchangeably.

Additional terms which are linked to the malicious behaviour of bots are *astroturfing*, *misdirection* and *smoke screening*. Ratkiewicz et al. (2011) describe political astroturf as "political campaigns disguised as spontaneous 'grassroots' behaviour that are in reality carried out by a single person or organisation. This is related to spam but with a more specific domain context, and potentially larger consequences" (p. 297). Astroturf is often linked to the intention to influence the opinion in a political debate and to create the impression that a vast majority is in favour of a certain position. In contrast to that, *smoke screening* entails the use of context-related hashtags on Twitter, to distract the readers from the main point of the debate (e.g. to use the hashtag *#syria* but talk about something unrelated to the war) (Abokhodair et al. 2015). *Misdirection* is similar to smoke screening, but goes a step further by using context-related hashtags without referring to the topic at all (e.g. use *#syria* but talk about something which is not related to Syria) (Abokhodair et al. 2015). All three forms of bot attacks can lead to a misconception of events and can influence e.g. the popularity of certain hashtags and topics in the related network. Subrahmanian et al. (2016) describe bots which seek to influence public discourse as "influence bots". Although influence bots have mainly been investigated in the context of political debates, it is easy to see how they could be used in a commercial context. The spreading of negative opinions about a certain product, brand or service can lead to a distorted perception of that product in the public opinion and a lasting damage for the enterprise. Whether in a political or a commercial context, as soon as bots aim at distorting the public perception, ethical concerns arise.

Finally, there are also bots that merely *are* – without being outright malicious or benign. The humoristic Twitter bots, as described by Veale et al. (2015), exemplify this type. Their only aim is to create funny and linguistically correct posts. Veale et al. identified two generations in the development of linguistic bots. While those of the first generation only made use of superficial language resources and did not manipulate text on a semantic level, second-generation bots apply semantic techniques and theories. Therefore, second-generation bots are linguistically more difficult to distinguish from human users. For these bots, we propose a third, *neutral* level for the intent, between malicious and benign.

## 4.2 Imitation of human behaviour

Besides the distinction between malicious and benign bots, another commonly made distinction is between bots that mimic human behaviour and those that do not (Boshmaf et al. 2013). Bots which pretend to be human users are often referred to as *social bots* (Boshmaf et al. 2011; He et al. 2017; Hegelich and Janetzko 2016; Igawa et al. 2016; Stieglitz et al. 2017). Abokhodair et al. (2015) describe social bots as automated social actors, which differ in their social skills and their intentions. They consider those automated social actors as dangerous which aim at being recognised as humans or companies, and which highlight one point of view to establish the impression of uniformity of opinions. Some of the previously described benign bots imitate human behaviour to some degree, too. For example, the humoristic bots described by Veale et al. (2015) produce messages which are very similar to content produced by human users. However, those bots do not try to hide that they are based on an algorithm. Often the profile description even states clearly that they are bots. As the boundaries between human and non-human behaviour are fading in some cases, some authors introduced the term *cyborg* to describe accounts which cannot be clearly categorised. These accounts can be humans who make use of automation techniques or bots which are managed by human beings (Chu et al. 2010). As social bots become more and more sophisticated, constantly advancing their cover-up techniques, their detection becomes increasingly challenging (Chavoshi et al. 2017; Everett et al. 2016). Researchers who programmed social bots report that social network operators were slow to identify and remove their bots (Boshmaf et al. 2011; Freitas et al. 2015).

As mentioned before, not all bots use sophisticated strategies in order to appear human. This is also true for accounts referred to as malicious. *Spam bots* often publish a large number of nearly identical messages in a short time. To the human observer, it is immediately clear that they are bots. Additionally, some bots that attempt to engage and converse with humans are based on simple rules. For example, on Twitter they might respond to all tweets that mention a keyword out of a predefined set by tweeting a generic response (Salto Martínez and Jacques García 2012). Moon (2017) also describes several accounts that do not attempt to hide that they are bots. The above considerations directly lead to the following categorisation of bot accounts on social media.

As we have shown, there are many different types of bots. For researchers, it might be quite helpful to distinguish between different classes of bots on social media because e.g. they have a different impact on communication on social media or because they require different approaches to be identified.

Also, as we have mentioned before, the terms used to describe different types of bots in the social media context are often imprecise – especially in the context of social bots. We argue that not every bot on the social media is a *social* bot, and that the term *social media* does not automatically imply that every automated (bot) account on such a platform is by definition a *social* bot. Instead, the term *social* refers to the imitation of human behaviour and the act of pretending to be a human with whom a social interaction is possible, "to act in ways that are similar to how a person might act in the social space" (Abokhodair et al. 2015, p. 840). Thus, we propose *imitation of human behaviour* to be the second dimension to discriminate between different kinds of bots on social media. Social bots are those bots that attempt to imitate humans to a large degree, while in contrast, a mere spam bot which only uses social media to disseminate a lot of messages exhibits a low degree of imitation.

These proposed dimensions are combined in the scheme shown in Table 1, which provides a way to organise different types of bots on social media. This categorisation covers accounts on social media sites that are controlled by bots, but differ regarding their intent and the degree to which they imitate human behaviour. Examples of social bots can be seen in the first row. They are social in the sense that they imitate human users to a high degree by writing original messages, sending friend requests, and sharing or retweeting information by other users. This definition is not limited to harmful accounts because helpful accounts can be as social, or even more social, than harmful ones. As we pointed out earlier, most definitions of social bots reflect these circumstances.

|  |  | Intent (Ferrara et al. 2016) | | |
| --- | --- | --- | --- | --- |
|  |  | Malicious | Neutral | Benign |
| Imitation of human behaviour (Boshmaf et al. 2013) | High: *Social* bots | • Astroturfing bot (Ratkiewicz et al. 2011)<br>• Social botnets in political conflicts (Abokhodair et al. 2015)<br>• Infiltration of an organisation (Elyashar et al. 2015)<br>• Influence bots (Subrahmanian et al. 2016)<br>• Sybils (Alarifi et al. 2016; Goga et al. 2015)<br>• Doppelgänger bots (Goga et al. 2015) | • Humoristic bots (Veale et al. 2015) | • Chat bots (Salto Martínez & Jacques García 2012) |
|  | Low to none | • Spam bots (Wang 2010)<br>• Fake accounts used for botnet command & control (Sebastian et al. 2014)<br>• Pay bots (Subrahmanian et al. 2016) | • Nonsense bots (Wilkie et al. 2015) | • News bots (Lokot & Diakopoulos 2016)<br>• Recruitment bots (Flores-Saviaga et al. 2016)<br>• Public Dissemination Account (Yin et al. 2014)<br>• Earthquake warning bots (Haustein et al. 2016)<br>• Editing Bots, Anti-Vandalism Bots on Wikipedia (Tsvetkova et al. 2017) |

*Table 1. Categorisation scheme of social media bot accounts*

## 5   Conclusion

This paper provides a comprehensive literature analysis on a very new topic that is becoming increasingly important for research and practice. Based on the insights from the review, a categorisation scheme was developed that includes and differentiates bots in social media on two dimensions: the intent (Ferrara et al. 2016) and the imitation of human behaviour (Boshmaf et al. 2013). This paper is the first to combine these two dimensions into a six-category system. For a coherent categorisation, the dimension of 'imitation of human behaviour' is divided into *high* and *low to none* while the dimension 'intent' includes *benign*, *neutral,* and *malicious*. By this means all social media bots, which have been analysed and discussed in the literature review could be classified. We follow Boshmaf et al. (2013) in defining *social* bots as bots which imitate human behaviour to a high degree, and give examples of such behaviour found in the literature. The literature review shows that the majority of papers on bots on social media address malicious bots. Social bots with a neutral or benign intent are an exception and are researched rarely. We do not assert that the proposed categorisation reflects the absolute truth, or is the only way to bring structure into the diversity of bots on social media. However, in our opinion it is a first helpful step for researchers and practitioners to categorise bots on social media.

Therefore, the first contribution of our article, which results from the systematic literature review, is to make this unstructured and heterogeneous research field more accessible. This article offers researchers an overview which will be especially helpful for those academics and practitioners who start investigating the phenomenon of social bots. Furthermore, researchers who already are engaged with the topic benefit from the categorisation, as it facilitates the localisation of the scope of their research in that field. Second, our categorisation system contributes to the research field by allowing bots to be assessed and analysed faster. Finally, we pointed out what separates *social* bots from other types of bots, leading to a more unified understanding of the phenomenon which can serve as the starting point for further analyses.

In further investigations, we plan to test the presented categorisation system empirically, to further prove its practical applicability. Further research may also examine which types of bots exist in different domains of social media communication. While, for example, harmful human-like bots that seek to influence human behaviour might be more frequent in politics than in sports and art, humorous bots might be prevalent in entertainment but not involved in a discussion about a current crisis situation. In addition to that, further research in this area could identify and improve effective methods for identifying bots. In this sense, the categorisation scheme raises the question whether different techniques are more helpful for certain bot categories. It can also be tested if the comparatively low number of papers researching potentially helpful and neutral bots which mimic human behaviour is due to this type of bot actually being rare, or if there is simply less research interest in bots which do not potentially harm social media users.

Overall, in this paper we pointed out a way to structure research on bots in social media and contributed to a broader understanding of this topic, thus providing a foundation for a more focused approach for future research.

**Acknowledgements:** This work was in part supported by the Deutsche Forschungsgemeinschaft (DFG) under grant No. GRK 2167, Research Training Group "User-Centred Social Media".